\documentclass[%
 reprint,
 amsmath,amssymb,
 aps,
prl,
]{revtex4-2}

\usepackage{physics}

\usepackage{graphicx}
\usepackage{dcolumn}
\usepackage{bm}
\usepackage[colorlinks=true, linkcolor=black, citecolor=black, urlcolor=black]{hyperref}

\usepackage{xcolor}
\usepackage{float}
\usepackage{natbib}
\begin{document}

\title{Environment induced dynamical quantum phase transitions in two-qubit Rabi model}

\author{G. Di Bello$^{1*}$}\author{A. Ponticelli$^{1}$}\author{F. Pavan$^{1}$}\author{V. Cataudella$^{2,3}$}\author{G. De Filippis$^{2,3}$}\author{A. de Candia$^{2,3}$}\author{C. A. Perroni$^{2,3}$} 
\affiliation{$^{1}$Dip. di Fisica E. Pancini - Università di Napoli Federico II - I-80126 Napoli, Italy}
\affiliation{$^{2}$SPIN-CNR and Dip. di Fisica E. Pancini - Università di Napoli Federico II - I-80126 Napoli, Italy
}
\affiliation{$^{3}$INFN, Sezione di Napoli - Complesso Universitario di Monte S. Angelo - I-80126 Napoli, Italy}

\begin{abstract}
The physics of quantum states beyond thermodynamic equilibrium represents a fascinating and cutting-edge research. Using numerical state-of-the-art approaches, we observe dynamical quantum phase transitions in the dissipative two-qubit Rabi model. By quenching the qubits-oscillator coupling, the system (Rabi + Environment) exhibits dynamical quantum phase transitions signalled by kinks of Loschmidt echo's rate function at parameter values close to thermodynamic transition. Notably, these transitions also manifest in two-qubit entanglement. While at equilibrium one class of Beretzinski-Kosterlitz-Thouless-type transitions occurs, non-equilibrium conditions reveal two classes of dynamical critical phenomena, depending on qubits' interactions and entanglement. When qubits directly interact, the kink critical exponent describes a linear behavior, reminiscent of nearest neighbors Ising chains, with short-range interactions dominating at short times. Conversely, non-interacting qubits exhibit critical exponents much smaller than unity due to bath-induced long-range interactions. These findings shed light on the complex behavior of dynamical quantum phase transitions in non-integrable models, showing unusual entanglement features and the environment's significant role.
\end{abstract}

\maketitle

One of the most challenging open problems in modern physics is the characterization of the transitions between different quantum phases. In recent decades, the importance of quantum phase transitions (QPTs) has grown considerably in various respects \cite{sachdev_2011}.
In general, QPTs are studied by examining long-term dynamics, asymptotic behavior of observables, or the non-analytical behavior of thermodynamic observables and correlators, which can be either local or non-local. The theory of QPTs offers insight into properties at thermodynamic equilibrium, leading to extensive research on phase transitions in quantum many-body systems far from equilibrium \cite{Schirò,Sciolla}. 
The fundamental concept of manipulating a control parameter to induce transitions between different phases has also extended to include open systems \cite{hwang2018dissipative,de2023signatures}.

Recently, both theoretical studies and experimental observations have focused on a different type of quantum phase transitions that occur on intermediate time scales, known as dynamical quantum phase transitions (DQPTs) \cite{Heyl_PhysRevLett.110.135704,heyl2018dynamical,heyl2019dynamical,Zvyagin}. In this context, specific quantities exhibit non-analytical behavior over time, where time itself acts as the primary driving parameter for this transition. DQPTs have been analyzed in various closed models, showing the emergence of singularities of the Loschmidt echo at the thermodynamic limit \cite{heyl2018dynamical}. In general, it is not guaranteed that an equilibrium QPT gives rise to a related non-equilibrium DQPT if the quench of some model parameters is made at initial time. More recently, DQPTs have been also found in models with only a few degrees of freedom, such as the closed quantum Rabi model when the oscillator frequency approaches zero \cite{puebla2020finite}.

At the same time, DQPTs have been also investigated in open systems, where the role of the Loschmidt echo is played by the fidelity of the subsystem density matrix operator at finite time with that at initial time \cite{link2020dynamical,nava2023lindblad,nava2024dissipation}. Also for such open systems DQPTs have been found in models with many degrees of freedom in the thermodynamic limit \cite{link2020dynamical} and with a few degrees of freedom with respect to different quench parameters \cite{dolgitzer2021dynamical}. In both cases the environment is treated at weak-coupling Lindblad level and does not actively trigger the phase transition, that is the QPT is already present in the closed system. For such complex systems, additional and more detailed investigations, making use of concepts from quantum information theory \cite{nielsen2010quantum,di2024optimal,osterloh2002scaling} are imperative to discover the active meaning of entanglement in both QPTs \cite{canovi2014dynamics} and DQPTs \cite{de2021entanglement}.

Recently, a study of the universality laws for DQPT has been proposed \cite{heyl2015scaling}, focusing on computing the critical exponent of the Loschmidt echo rate function. For nearest neighbors Ising chains, a linear behavior with a critical exponent of 1 has been observed. It is known that long-range interactions characterized by a $1/r^{\gamma}$ behavior ($\gamma\leq 2$) can lead to a DQPT \cite{vzunkovivc2018dynamical}, although the precise value of the critical exponent remains unclear. More importantly, in long-range interacting Ising models, the various observed DQPTs are not directly related to the underlying equilibrium phase transition \cite{heyl2018dynamical}. Furthermore, other studies \cite{trapin2021unconventional,wu2020dynamical} have demonstrated that introducing random interactions in an Ising model or tuning the quench parameters can lead to a reduction of the critical exponent from its expected integer value of 1 to small non-integer values.

\begin{figure}[htbp]
    \begin{center}
        \includegraphics[scale=0.39867]{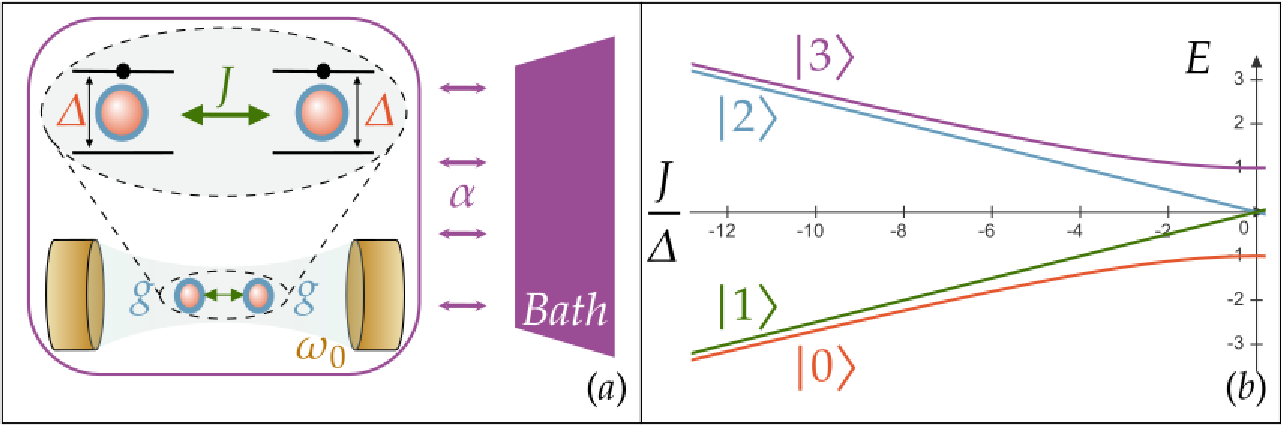}
     
        \caption{\label{fig:modspettro}\textit{Two-qubit dissipative Rabi model and closed system energy diagram}\\
        Model described by Hamiltonians \ref{eq:hsyst} and \ref{eq:hbath} ($a$): two qubits with energy gap $\Delta$ interacting through $J$ and connected to an oscillator through $g$. The cavity housing the qubits is in contact with an Ohmic bath through $\alpha$. Energy spectrum (in units of $\Delta$) of the system consisting of the two interacting qubits ($b$), as a function of the ratio $J/\Delta$ in the ferromagnetic region (negative values of $\frac{J}{\Delta}$ on x-axis). } 
    \end{center}
\end{figure}

In this Article, we analyze a dissipative two-qubit Rabi model, which involves two interacting qubits coupled to an oscillator, which in turn interacts with an Ohmic bath (see Figure \ref{fig:modspettro}.a). We emphasize that, differently from \cite{grimaudo2023quantum}, the model under consideration here explicitly incorporates dissipation \cite{prlnote}. Since the focus is on dynamical transitions induced by the environment, we consider the bath and the two-qubit-oscillator system on the same footing within all the coupling regimes, overcoming the limitations of the perturbative Lindblad approach \cite{di2023qubit}. Since an analytical solution does not exist to compute the time-dependent system behavior and the Loschmidt echo for the entire system, including the multiple bath degrees of freedom, we use the time-dependent variational principle (TDVP) with the matrix product state (MPS) ansatz \cite{gnote}, making a quench on the qubits-oscillator coupling constant. As we will discuss in the following, we find evidences of DQPTs from the singularities of the related rate function \cite{hnote}.

In order to characterize the nature of these transitions, in this Article, we first analyze the thermodynamic properties of the system, employing the worldline quantum Monte Carlo (WLMC) method, which is useful to compute mean values and correlation functions of the observables at imaginary times. Moreover, density-matrix renormalization group (DMRG) is useful to compute the ground state of the Hamiltonian as an MPS \cite{gnote}. In addition to the previously identified signatures detailed in Ref. \cite{de2023signatures}, our study fully characterizes this QPT through the bimodal nature of the magnetization distribution and the entanglement properties of qubits, such as von Neumann entropy and concurrence. This analysis pinpoints that the DQPTs mirror QPT features in non-equilibrium conditions. Singularities in the rate function begin to emerge, exhibiting similarities with integrable systems \cite{Heyl_PhysRevLett.110.135704,heyl2018dynamical,heyl2019dynamical,Zvyagin}, particularly for values of qubit-oscillator coupling that are close to those where also QPT occurs. Additionally, we explore the universality laws of DQPTs, examining the critical exponent of the rate function of the Loschmidt echo. We find two classes of universality, in contrast with the case of the equilibrium QPT. We indeed analyze both cases: when the two qubits ferromagnetically interact (covered in the main text) and when they do not (discussed in the Supplemental Material \cite{gnote}), finding significant differences in their behaviors related to the presence of interactions and entanglement. 

\textbf{Dissipative quantum Rabi model with two interacting qubits} 
We consider two interacting qubits connected through a harmonic oscillator to an Ohmic bath \cite{di2024optimal,bagninote} (see Figure \ref{fig:modspettro}.a). We set $\hbar=k_B=1$. The Hamiltonian that describes the system is given by: $H=H_{S} + H_{B} + H_{S-B}$. Here, the system (qubits and oscillator) energy $H_{S}$ is defined as:
\begin{equation}\label{eq:hsyst}
    H_{S} = -\frac{\Delta}{2}(\sigma_x^1 + \sigma_x^2) + \frac{J}{4}\sigma_z^1\sigma_z^2 + \omega_0 a^{\dagger}a + g(a+a^{\dagger})(\sigma_z^1 + \sigma_z^2),
\end{equation}
where $\Delta$ is the frequency of the two qubits, $J$ is the strength of the interaction between them, and $\sigma_i^j$ (with $i=x,y,z$ and $j=1,2$) are the Pauli matrices. The oscillator frequency is $\omega_0$, and $a$ ($a^{\dagger}$) are the annihilation (creation) operators. The parameter $g$ represents the coupling strength between the qubits and the oscillator. The bath Hamiltonian and its interaction with the system are given by:
\begin{equation}\label{eq:hbath}
    H_{B} + H_{S-B} = \sum_{i=1}^{N} \left[\frac{p_i^2}{2M_i} + \frac{k_i}{2}(x_0 - x_i)^2\right].
\end{equation}
The bath is represented as a collection of $N$ oscillators with frequencies $\omega_i = \sqrt{\frac{k_i}{M_i}}$, and coordinates and momenta are denoted by $x_i$ and $p_i$, respectively. Additionally, $x_0$ denotes the position operator of the resonator with mass $m$: $x_0 = \sqrt{\frac{1}{2m\omega_0}}(a + a^{\dagger})$. This interaction with the bath describes dissipation as proposed by Caldeira-Leggett \cite{CaldeiraLeggett81,weiss2012quantum} and can be experimentally realized by ultrastrongly coupling a flux qubit to its resonator, which is further coupled to an Ohmic bath \cite{magazzu2019transmission,goorden2004entanglement,malekakhlagh2019quantum,cai2021observation}. The coupling to the bath induces renormalization effects on several parameters: the oscillator frequency $\Bar{\omega}_0$, as well as the interaction strengths $\Bar{g}$ and $\Bar{g_i}$ (further detailed in Supplemental Material \cite{gnote}). This results in the following bath spectral density $J(\omega)= \sum_{i=1}^N \frac{k_i\omega_i}{4m\Bar{\omega}_0} \delta(\omega-\omega_i) = \frac{\alpha}{2}\omega f\left(\frac{\omega}{\omega_c}\right)$ where $\alpha$ controls the system-bath coupling. Here $f\left(\frac{\omega}{\omega_c}\right)$ is a function that depends on the cutoff frequency for the bath modes, $\omega_c$, which governs the behavior of the spectral density at high frequencies. This function is taken as $f\left(\frac{\omega}{\omega_c}\right)=\Theta\left(\frac{\omega_c}{\omega}-1\right)$, where $\Theta(x)$ is the Heaviside step function. The cutoff frequency is typically chosen to be the largest energy scale in the system. In the following we set: $\omega_0=\Delta,\,J=-10\Delta$ (ferromagnetic interaction)$,\,\alpha=0.1,\,\omega_c=30\Delta$.\\
It is worth noticing that the system can be mapped to an equivalent model of two interacting qubits in contact, through $\sigma_z$, with a structured bath whose spectral density shows a peak at the oscillator frequency \cite{zueco2009qubit,de2023signatures,gnote}. The spectrum of the two-qubit Hamiltonian, $H_{qub}=H_{S}(g=0)$, is shown in Figure~\ref{fig:modspettro}.b.

\textbf{QPT evidences at thermodynamic equilibrium} 

We first investigate the equilibrium properties of the system (1.-2.), using two different approaches. The first method is the WLMC approach, which is based on path integrals \cite{de2020quantum,de2021quantum,de2023signatures,gnote}. Here, the structured bath degrees of freedom are eliminated, resulting in an effective Euclidean action \cite{weiss2012quantum,winter2009quantum,feynman1950mathematical,de2023signatures}, with the kernel expressed in terms of the structured spectral density $J(\omega)$. This structure is characteristic of a spin-boson model but is now extended to involve two qubits interacting with the bath, as outlined in \cite{magazzu2019transmission, zueco2009qubit, gnote}. This transformation leads to a classical system of spin variables distributed along two chains, each of them with length $\beta=1/T$. The spins experience long-range ferromagnetic interactions both within each chain and between the two chains. The functional integral is computed using a Poissonian measure and adopts a hybrid algorithm \cite{rieger1999application,winter2009quantum}, based on an alternation of Wolff’s \cite{wolff} and Metropolis moves. Notably, as $\omega_0$ remains constant and $\beta$ tends toward infinity, the kernel displays a power asymptotic behavior, $K(\tau)\simeq 1/\tau^2$. This power-law behavior with an exponent of $2$ determines the onset of a Beretzinski-Kosterlitz-Thouless (BKT) QPT.\\
The second method we employ is the DMRG \cite{gnote}, an adaptive algorithm for computing eigenstates of many-body Hamiltonians. It is particularly effective for calculating low-energy properties of one-dimensional and two-dimensional quantum systems. DMRG uses the MPS representation to determine the ground state of low-dimensional quantum systems. In particular, we use the ITensor library \cite{fishman2022itensor} for a system with a bath of $N=300$ harmonic oscillators.\\

\begin{figure}[H]
    \begin{center}
        \includegraphics[scale=0.22]{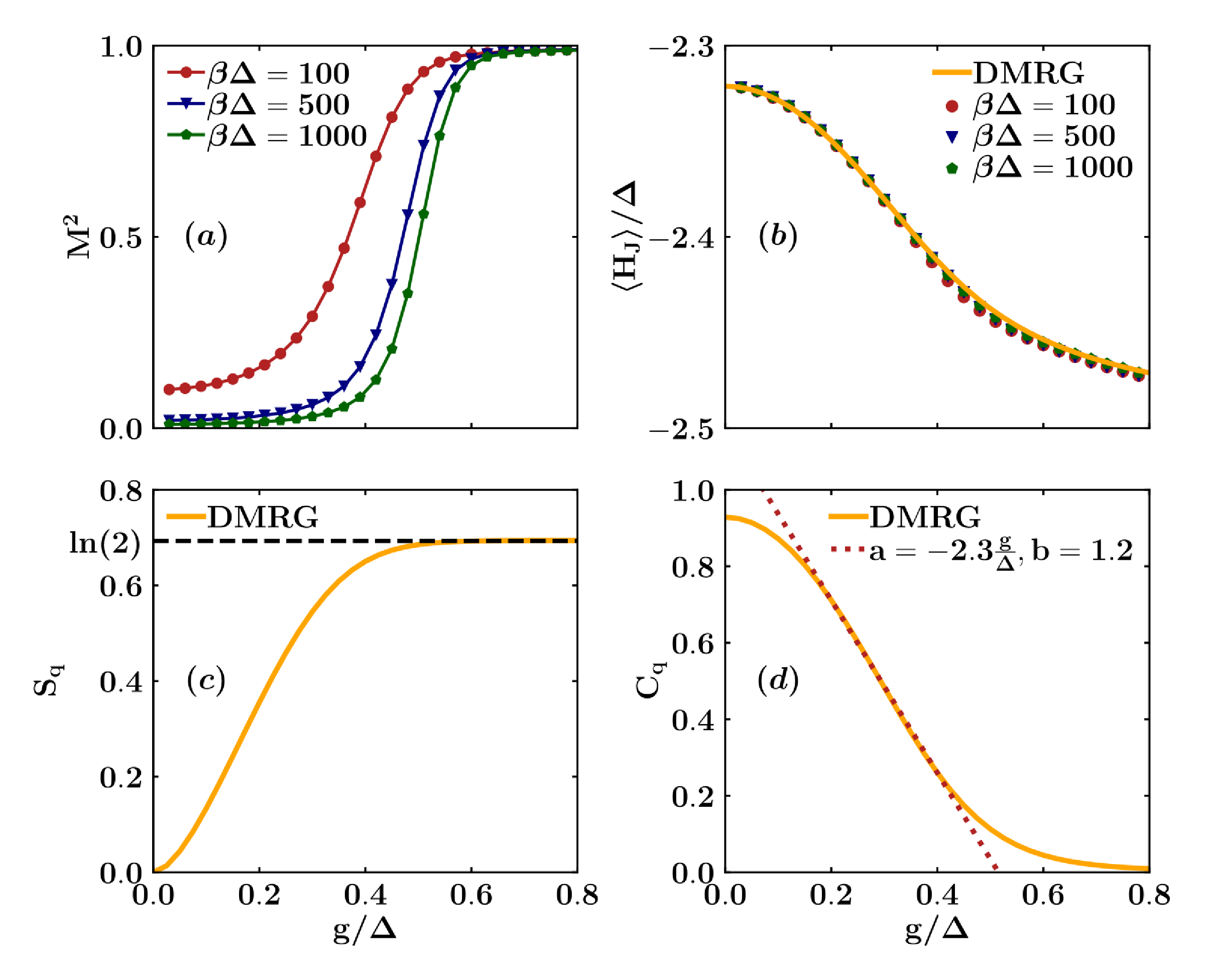}
     
        \caption{\label{fig:te1}\textit{Order parameter crossover and entanglement signatures of QPT at thermodynamic equilibrium}\\
        Qubits' squared magnetization $M^2$ ($a$), interaction energy between the qubits $\langle H_J\rangle/\Delta$ ($b$), von Neumann entropy $S_q$ ($c$) and concurrence $C_q$ ($d$) as functions of $g/\Delta$, computed through WLMC and DMRG. For the WLMC method the calculations are made for $\beta\Delta \in[100,1000]$.} 
    \end{center}
\end{figure}

\textbf{Squared magnetization, interaction energy, and entanglement.} In Figure \ref{fig:te1}.a, we present the squared magnetization of the qubits, defined as $M^2 = \frac{1}{4\beta}\int_0^{\beta} d \tau \langle(\sigma_z^1+\sigma_z^2)(\tau)(\sigma_z^1+\sigma_z^2)(0)\rangle$, where $\tau$ labels the positions of the spins on the two equivalent chains (corresponding to the two spins labeled by the superscripts $^1$ and $^2$), after eliminating the bath degrees of freedom in favour of an effective classical system of spins as discussed in \cite{gnote,magazzu2019transmission}. We plot this magnetization as a function of $g/\Delta$ for three different inverse temperatures, $\beta\Delta=\{100,500,1000\}$. The data obtained using WLMC method exhibit a crossover for $M^2$ from $0$ to $1$ increasing $g/\Delta$ that becomes sharper and sharper lowering the temperature. This behavior suggests the occurrence of the BKT QPT, which is estimated to set in at a critical value of $g_c \approx 0.5\Delta$ from $\beta\Delta=10^3$ data. Additionally, we calculate the mean value of the interaction Hamiltonian between the two qubits, denoted as $\langle H_J\rangle = J\langle\sigma_z^1\sigma_z^2\rangle/4$, as a function of $g/\Delta$ for the same three temperature values (see Figure \ref{fig:te1}.b). We emphasize that both the WLMC and DMRG methods yield consistent results. Note that the interaction becomes more negative with increasing values of $g$ since the bath induces an effective ferromagnetic coupling between the spins.\\
\noindent
To analyze the entanglement properties of the system in the presence of the QPT, we determine the qubits' density operator $\rho_q$ and compute the qubits' entropy, denoted as $S_{q}=-\text{Tr}\{\rho_q \ln{\rho_q}\}$, and the concurrence, given by $C_{q}=\max(0,\lambda_1-\lambda_2-\lambda_3-\lambda_4)$. Here, $\lambda_i$ represents the eigenvalues of the Hermitian matrix $\sqrt{\sqrt{\rho_q}\Tilde{\rho_q}\sqrt{\rho_q}}$, and $\Tilde{\rho_q}=(\sigma_y^1\otimes\sigma_y^2)\rho_q^*(\sigma_y^1\otimes\sigma_y^2)$, with $^*$ indicating complex conjugation. In Figures \ref{fig:te1}.c and \ref{fig:te1}.d, we present the entropy and entanglement as functions of $g/\Delta$, computed using the DMRG algorithm. The entropy increases for values of $g$ near the critical point, approaching a value of approximately $\ln(2)$ just at the critical values determined by the WLMC approach. We also notice the similarity between Figures \ref{fig:te1}.a and \ref{fig:te1}.c. 
In contrast, as shown in Figure \ref{fig:te1}.d, the concurrence decreases as a function of $g$, approaching zero. Moreover, through a linear fit of the concurrence within the critical region ($C_q=a g/\Delta+b$ in Figure \ref{fig:te1}.d), it becomes evident that the line intersects the $g$-axis at approximately $g\approx 0.5\Delta$, a value close to the estimated critical point. This behavior can be explained by the qubits approaching a two-degenerate state at the BKT QPT, where they are either both in the ``up" or ``down" state. This clearly results from a lack of entanglement between the two qubits and an enhancement of entanglement of each of them with the bath \cite{de2021quantum}.\\
\begin{figure}[H]
    \begin{center}
        \includegraphics[scale=0.22]{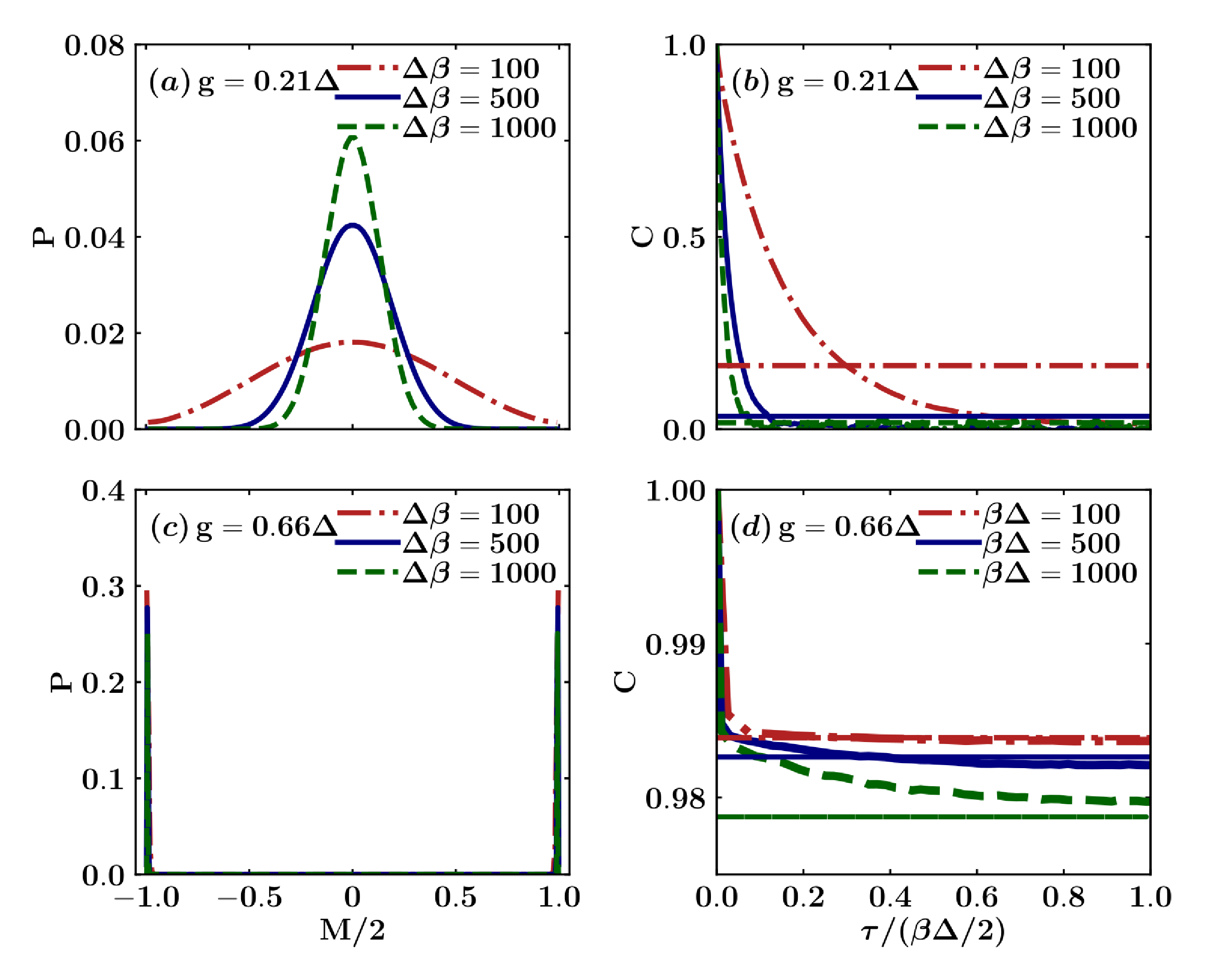}
     
        \caption{\label{fig:distrcorr}\textit{Bimodal magnetization distribution and correlations signal the QPT at thermodynamic equilibrium}\\
        Qubits' magnetization distribution $P(M/2)$ ($a$) and normalized correlation function $C=C(\tau/(\beta\Delta/2))/C(0)$ ($b$). We consider two scenarios: one with $g=0.21\Delta$ before the transition and another with $g=0.66 \Delta$ after the transition. These calculations are performed using the WLMC method at inverse temperatures of $\beta\Delta \in [100,1000]$.} 
    \end{center}
\end{figure}
\textbf{Magnetization distribution and correlations.} Since $M^2$ displays a crossover from $0$ to $1$ (Figure \ref{fig:te1}.a), one naturally wonders if this is related to the onset of a BKT QPT. This question can be better addressed studying the distribution of the normalized magnetization, denoted as $P(M/2)$. In Figure \ref{fig:distrcorr}, we plot the magnetization distribution for two different values of $g/\Delta$, smaller and larger of the estimated one for BKT transition. When $g=0.21\Delta<g_c$ (Figure \ref{fig:distrcorr}.a), the distribution exhibits a single peak centered at $M = 0$. On the other hand, for $g=0.66\Delta > g_c$ (Figure \ref{fig:distrcorr}.c), it acquires a bimodal character, with two peaks centered at $M/2 = \pm 1$, again with the same vanishing mean value. We emphasize that, above $g_c$, the distribution develops two peaks that are expected to become two delta functions, centered at $\pm \sqrt{M^2}$, in the thermodynamic limit. It's also worth noting that the formation of a bimodal distribution is clearly related to the crossover observed in $M^2$. This behavior, reminiscent of classical thermodynamics, signals the emergence of a QPT. In addition, we can examine the correlations $C(\tau)=\langle(\sigma_z^1+\sigma_z^2)(\tau)(\sigma_z^1+\sigma_z^2)(0)\rangle$ as another indicator of the occurrence of the QPT. Figure \ref{fig:distrcorr} shows the distinct behavior before and after the onset of the transition. Specifically, the normalized correlation function, defined as $C=C(\tau/(\beta/2))/C(0)$, tends toward $0$ as $\tau$ approaches $\beta/2$ before the critical point (Figure \ref{fig:distrcorr}.b) and converges to a finite value, i.e. $M^2$, after the transition (Figure \ref{fig:distrcorr}.d), indicating the long-range nature of the correlations between the spins above $g_c$.

\textbf{Dynamics of energy and entanglement} 

\begin{figure}[H]
    \begin{center}
        \includegraphics[scale=0.22]{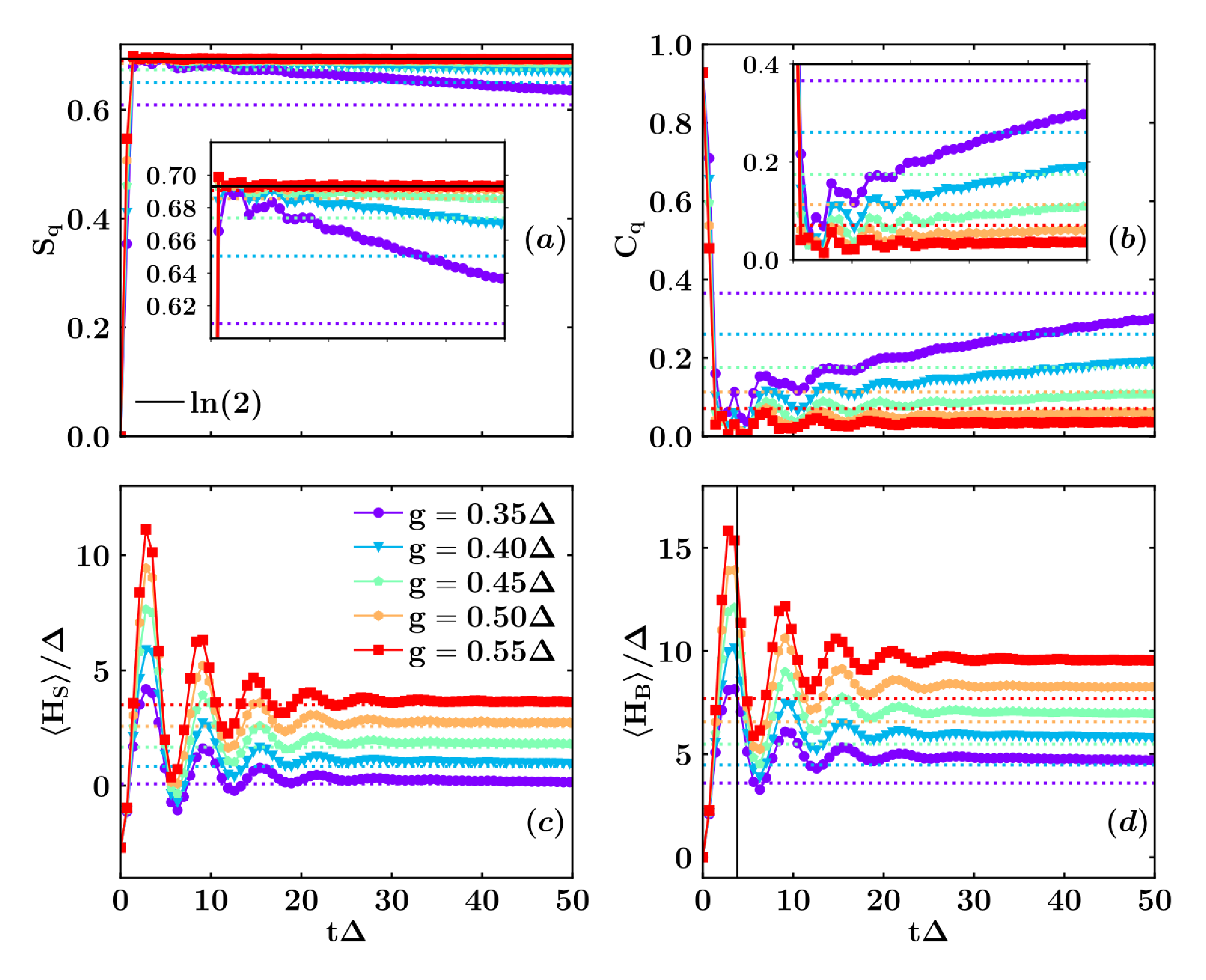}
     
        \caption{\label{fig:enent}\textit{Energy and entanglement dynamical behavior}\\
        Qubits' entropy $S_q$ ($a$) and concurrence $C_q$ ($b$), system energy $\langle H_S(t)\rangle/\Delta$ ($c$) and bath energy $\langle H_B(t)\rangle/\Delta$ ($d$) as functions of dimensionless time $t\Delta $ for different values of the coupling $g\in[0.35,0.55]\Delta$, crossing the critical point, calculated through TDVP and DMRG.} 
    \end{center}
\end{figure}

We investigate the out-of-equilibrium properties of the system, focusing on the behavior of energy and entanglement over time. To accomplish this, we employ the TDVP algorithm \cite{haegeman2011time,haegeman2016unifying,paeckel2019time,gnote}, implemented using the ITensor library \cite{fishman2022itensor}, to evolve the wavefunction of the entire system represented as an MPS. The adoption of this technique proves advantageous for our system strongly coupled to the environment, enabling us to achieve long simulation times. Consequently, we can compare these behaviors with those computed using the DMRG at thermodynamic equilibrium. Specifically, we choose the ground state of the Hamiltonian $H_S (g = 0)$ (state $\ket{0}$ in Figure \ref{fig:modspettro}) as the initial state for simulations and calculate the qubits' von Neumann entropy $S_q(t)$, the concurrence $C_q(t)$, and the mean values of the various contributions to the total energy of the system, including $\langle H_S(t)\rangle$, $\langle H_B(t)\rangle$, and $\langle H_{S-B}(t)\rangle$ for different values of $g$, crossing the critical point. Figures \ref{fig:enent}.a and \ref{fig:enent}.b demonstrate that both entropy and concurrence are approaching thermodynamic values. Moreover, the greater $g$, the less time the system needs to reach the equilibrium values. In the Supplemental Material we show that for $J=0$ these quantities reach their equilibrium values at very earlier times \cite{gnote}. 
Regarding energy contributions, as depicted in Figure \ref{fig:enent}.c, the mean value of the system's energy $\langle H_S(t)\rangle$ approaches its thermodynamic equilibrium value after approximately $t\Delta =30$, while the mean value of the bath energy $\langle H_B(t)\rangle$ (see Figure \ref{fig:enent}.d) never reaches the DMRG values. We also computed the mean value of the interaction energy $\langle H_{S-B}(t)\rangle$, although it is not shown here, which reaches thermodynamic equilibrium on the same time scale as the system. The total energy is conserved, the bath remains at zero temperature, but accumulating bosons, i.e. absorbing the energy difference relative to the ground-state energy calculated through DMRG. This phenomenon can also be understood in terms of quasiparticles. When the coupling $g$ is small, the system reaches the equilibrium of its Hamiltonian $H_S$. However, as $g$ increases, one can envision a new Hamiltonian involving non-interacting quasiparticles dressed by the bath bosons, resulting in a small residual interaction energy. Consequently, quasiparticles reach the thermodynamic equilibrium in the presence of additive free bosons that are not able to modify the bath temperature. After analyzing asymptotic times, in the following we will focus our attention on smaller time scales. 

\textbf{Dynamical quantum phase transitions: Loschmidt echo} 

\begin{figure}[htbp]
    \begin{center}
        \includegraphics[scale=0.22]{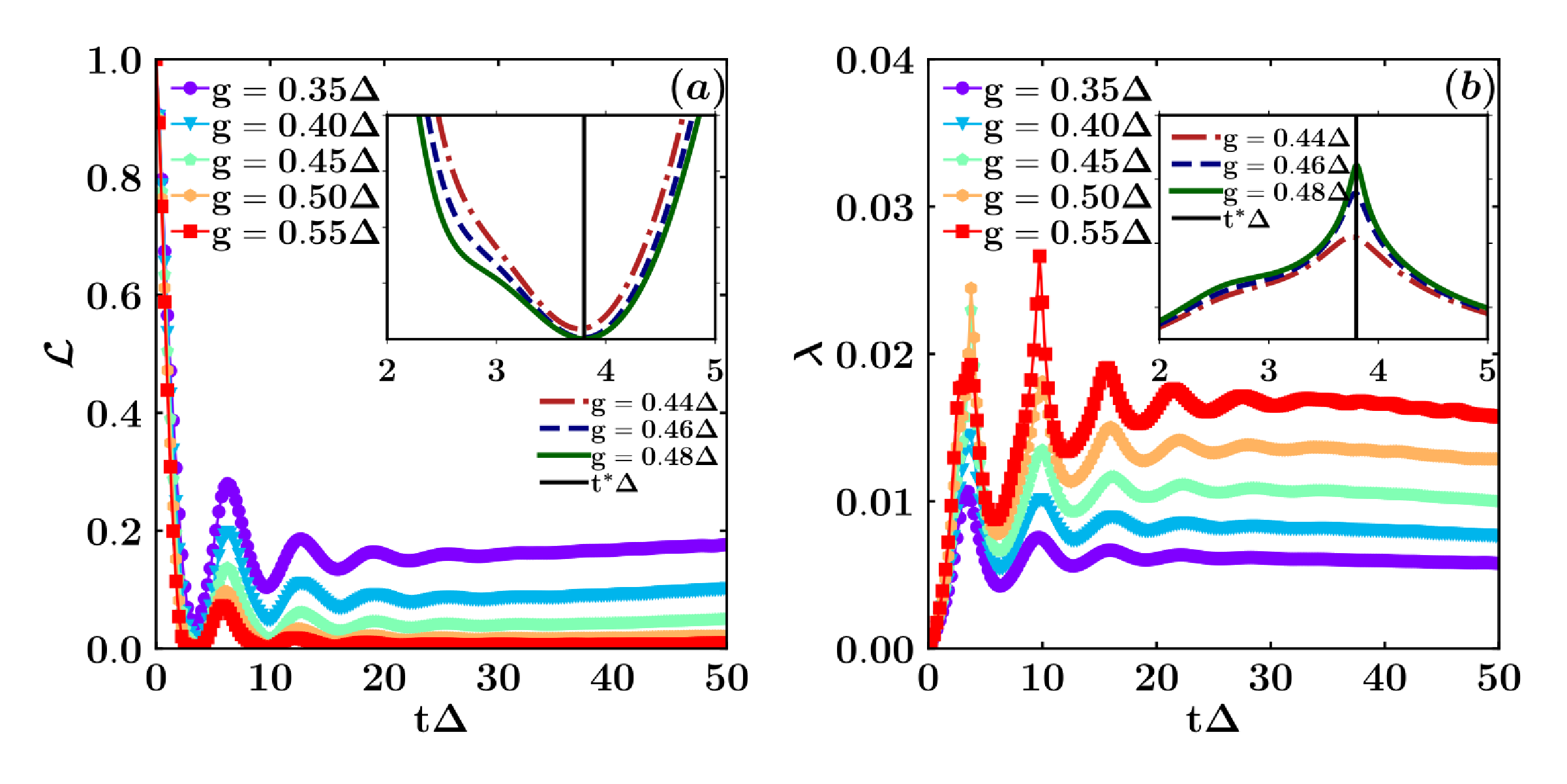}
     
        \caption{\label{fig:echo} \textit{Non-analytical behavior of Loschmidt echo's rate function signals DQPTs}\\
        Loschmidt echo $\mathcal{L}(t)$ ($a$) and rate function $\lambda(t)$ ($b$) as functions of dimensionless time $t\Delta $ for different values of the coupling $g\in[0.35,0.55]\Delta$, crossing the critical point, computed through TDVP. The insets provide a zoomed-in view near the transition, allowing identifying the critical time $t^*\Delta\approx3.8$.} 
    \end{center}
\end{figure}

It has been demonstrated in \cite{heyl2018dynamical,heyl2019dynamical} that a closed quantum many-body system can undergo a DQPT without any external control parameters, such as temperature or pressure. The typical non-analyticities of phase transitions manifest over time in the matrix element of the system unitary evolution operator on the initial state, i.e. the Loschmidt echo $\mathcal{L}(t)$ (see Figure \ref{fig:echo}.a). To observe such a phase transition, the procedure involves preparing the system in a well determined initial state inducing a quench in a parameter on which the Hamiltonian depends. Subsequently, the system evolves with the full Hamiltonian after the quench, and the Loschmidt echo is computed over time \cite{medvidovic2023variational}. It can be expressed as $\mathcal{L}(t)=e^{-N\lambda(t)}$ taking into account the exponential dependence on the system's degrees of freedom, denoted by $N$. Therefore, the rate function $\lambda(t)$ (see Figure \ref{fig:echo}.b) is the key property to monitor over time in order to observe non-analyticities. Formally, there exists an equivalence between the rate function and the free energy derived from a complex partition function, demonstrating the presence of singular points. Here, we aim to compute the same quantities in the closed system, which includes the bath. \\
Figure \ref{fig:echo} illustrates the echo and the rate function over time for different values of $g$ around the transition. Additionally, the inset clearly demonstrates how the scalar product between the evolved state and the initial one becomes zero (see inset of Figure \ref{fig:echo}.a) and the kink becomes narrower and higher as the critical point is approached at time $t^*\Delta \approx 3.8$ (see inset of Figure \ref{fig:echo}.b). In Figure \ref{fig:enent}.d we can see that the first maximum at short times occurs at around the critical time indicated by the black vertical line, as another marker of the transition in the dynamics. Beyond the critical region, the peak occurs at earlier times, and multiple peaks emerge over time in the rate function. This behavior occurs because the system can transition to the other phase more rapidly with higher couplings to the bath. The scalar product which vanishes indicates that the bath, including a lot of excited bosons, significantly differs from the initial vacuum state, leading to the orthogonality catastrophe, a characteristic feature of a phase transition. We emphasize that, through the quench, we are probing the first excited states and fluctuations are responsible for the observation of the DQPTs, reminiscent of the QPT occurring at thermodynamic equilibrium at zero temperature in the ground state of the entire Hamiltonian \cite{hnote}.

\begin{figure}[htbp]
    \begin{center}
        \includegraphics[scale=0.195]{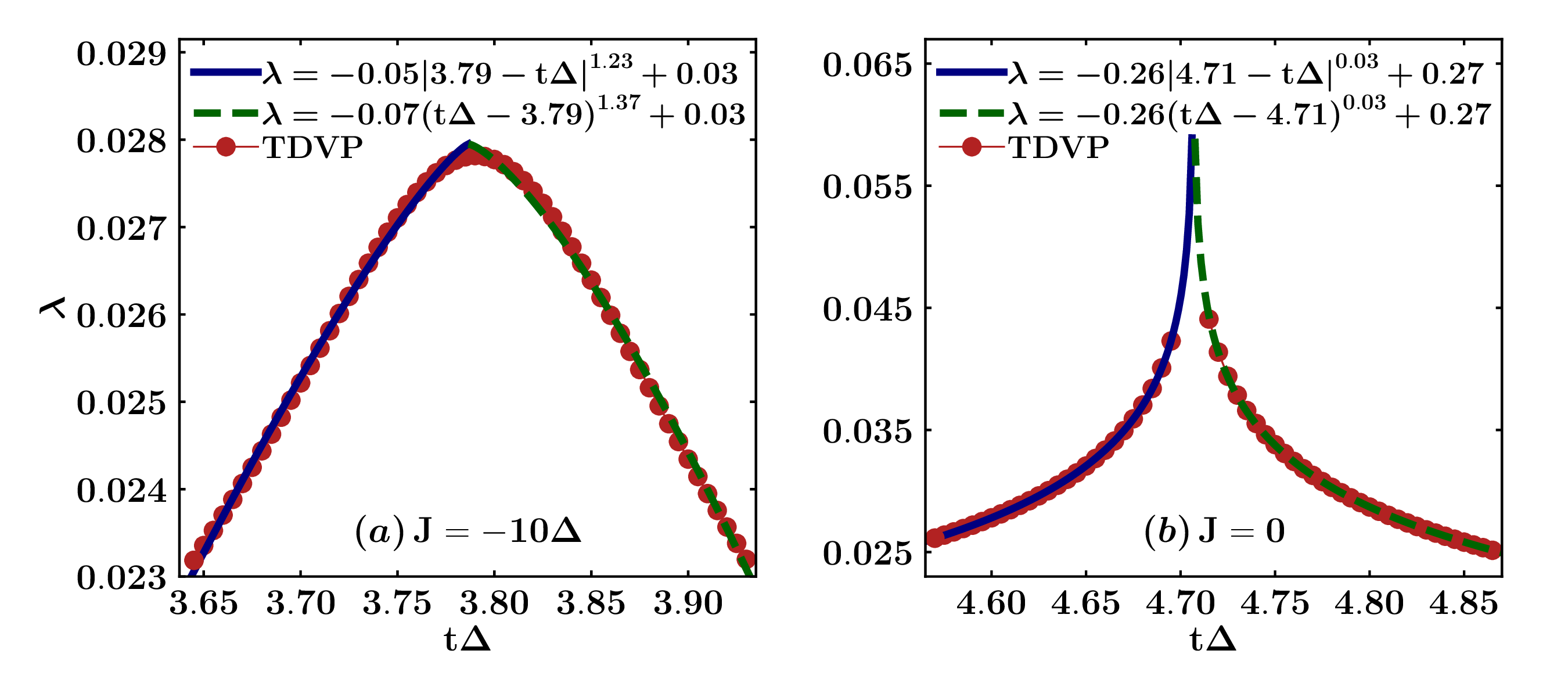}
     
        \caption{\label{fig:echofit}\textit{Critical exponent of non-analicities of Loschmidt echo's rate function to classify DQPTs}\\    
        Loschmidt echo rate function $\lambda(t)$ as functions of dimensionless time $t\Delta $ for the two cases $J=-10\Delta,\,g=0.48\Delta$ ($a$) and $J=0,\,g=0.58\Delta$ ($b$). The red points represent data computed through TDVP. For each case, we provide fitting functions for both the left branch (blue solid line) and the right branch (green dashed line). The legend indicates the parameter values obtained from the fit.} 
    \end{center}
\end{figure}

To further classify the DQPTs, we study the critical exponent of the rate function of the Loschmidt echo, focusing on the $J=-10\Delta$ case and the $J=0$ case. The corresponding results are presented in Figure \ref{fig:echofit}.a and Figure \ref{fig:echofit}.b, respectively. We fit the left branch of the data, preceding the peaks, using the function $\lambda(t)=a_L\left|d_L-t\right|^{b_L}+c_L$ with four free parameters. Subsequently, we fix $d_L$ and $c_L$, the coordinates of the peak, for the right branch fitting, which employs the function $\lambda(t)=a_R(t-d_L)^{b_R}+c_L$ with two free parameters \cite{fitnote}. 

Our analysis reveals that in the case of $J=-10\Delta$, the critical exponent $b$ is approximately 1 for both branches. However, in the $J=0$ case, the critical exponent $b$ exhibits significantly different behavior, being on the order of 0.03 for both branches. These distinct behaviors point out the significant impact of interactions and entanglement.

As explained earlier and detailed in the Supplemental Material \cite{gnote}, our model can be mapped by eliminating structured bath degrees of freedom. The resulting effective Euclidean action is characterized by a spin-boson model extended to involve two qubits interacting with the bath, yielding a classical system of spin variables distributed along chains of length $\beta$ with long-range ferromagnetic interactions. The additional interaction between the two qubits, governed by $J$, induces short-range interactions between the spin chains. In the case of $J=-10\Delta$, interactions between the two chains and initial qubits' entanglement significantly contribute. Short-range interactions play a crucial role at short times, altering the critical behavior compared to the $J=0$ case, where the behavior is primarily governed by long-range interactions. It is known that a critical exponent of 1 is expected for a nearest neighbors Ising chain, resembling our case $J=-10\Delta$, suggesting a similar linear behavior \cite{heyl2015scaling}. Moreover, in the case of $J=0$, only long-range ferromagnetic interactions are initially present along the chains, with no entanglement between qubits. Interactions between the chains are induced by their coupling to the bath, which reduces the critical exponent \cite{trapin2021unconventional, wu2020dynamical}.

\textbf{Conclusions} 

We have shown that an interacting two-qubit model strongly coupled to a $T=0$ bath exhibits dynamical phase transitions not present in the closed configuration (no coupling with the bath). We also provide evidence that the DQPT’s are related to a BKT equilibrium QPT. Our analysis not only encompasses the investigation of singularities in the rate function but also highlights clear signatures of the DQPT through entanglement properties, such as the qubits' von Neumann entropy and concurrence. Additionally, we observe the impact of interaction by examining the critical exponent of the singularities in the Loschmidt echo rate function for both the interacting ($J=-10\Delta$) and non-interacting ($J=0$) cases. We find two classes of emergent dynamical critical phenomena demonstrating a shift from exponent 1 in the former case to 0.03 in the latter, attributed to the different role of entanglement dynamics. Our findings challenge the conventional belief that the environment invariably exerts a detrimental influence on the DQPT of the system. Instead, we reveal that the environment can induce novel dynamical phase transitions. These results pave the way for experimental investigations. Indeed, recent advancements in quantum technologies, such as superconducting qubits and circuit QED setups, make feasible the experimental realization of our proposed model and the observation of dynamical quantum phase transitions induced by the environment, opening up new possibilities in quantum devices.

\textbf{Methods} 

Equilibrium properties are explored using two distinct methods. Firstly, the World Line Monte Carlo (WLMC) approach, based on path integrals, eliminates the structured bath degrees of freedom, yielding an effective Euclidean action. A cluster algorithm, combining Wolff's \cite{wolff} and Metropolis moves, is employed to approach the thermodynamic limit by gradually decreasing the temperature. The second approach, Density Matrix Renormalization Group (DMRG), utilizes an adaptive algorithm and Matrix Product State (MPS) representation to determine ground state properties in systems with a bath of $N = 300$ harmonic oscillators.

For out-of-equilibrium properties, the Time-Dependent Variational Principle (TDVP) algorithm is employed to evolve the wavefunction of the entire system, represented as an MPS. This technique facilitates long simulation times, even for systems with a bath of $N = 300$ harmonic oscillators, allowing for comparison of long-time stationary behaviors with those computed using DMRG at thermodynamic equilibrium.

Further details on both the equilibrium and out-of-equilibrium techniques are provided in the Supplemental Material \cite{gnote}.

\bibliography{biblio}

\begin{thebibliography}{49}
\bibitem [{fit()}]{fitnote}%
  \BibitemOpen
  \href@noop {} {}\bibinfo {note} {In the fit performed for the $J=0$ case, we exclude the central points next to the peak.}\BibitemShut {Stop}%
\end{thebibliography}

\textbf{Acknowledgements}

\begin{acknowledgements}
    G.D.F. acknowledges financial support from 376 PNRR MUR Project No. PE0000023-NQSTI. C.A.P. acknowledges founding from the European Union's Horizon Europe research and innovation programme under grant agreement n. 101115190. G.D.F. and C.A.P. acknowledge founding from the PRIN 2022 project 2022FLSPAJ ``Taming Noisy Quantum Dynamics" (TANQU). C.A.P. acknowledges founding from the PRIN 2022 PNRR project P2022SB73K - ``Superconductivity in KTaO3 Oxide-2DEG NAnodevices for Topological quantum Applications" (SONATA) financed by the European Union - Next Generation EU.
\end{acknowledgements} 

\textbf{Author contributions}

G. D. B. wrote the code and ran the numerical MPS simulations. A. P. and A. d. C. wrote the code and ran the numerical WLMC simulations. F. P., V. C., G. D. F., and C. A. P. developed the complete theoretical framework. All authors discussed the results and contributed to writing the manuscript.

\textbf{Supplementary information} accompanies this paper.

\textbf{Competing interests.} The authors declare no competing financial interests.
\end{document}


\begin{frontmatter}
\renewcommand{\theaffn}{\arabic{affn}}

\title{Supplemental Material: Environment induced dynamical quantum phase transitions in two-qubit Rabi model}

\author{G. Di Bello$^{1}$}\author{A. Ponticelli$^{1}$}\author{F. Pavan$^{1}$}\author{V. Cataudella$^{2,3}$}\author{G. De Filippis$^{2,3}$}\author{A. de Candia$^{2,3}$}\author{C. A. Perroni$^{2,3}$} 
\affiliation{Dip. di Fisica E. Pancini - Università di Napoli Federico II - I-80126 Napoli, Italy}
\affiliation{SPIN-CNR and Dip. di Fisica E. Pancini - Università di Napoli Federico II - I-80126 Napoli, Italy
}
\affiliation{INFN, Sezione di Napoli - Complesso Universitario di Monte S. Angelo - I-80126 Napoli, Italy}

\begin{abstract}
In this supplemental material, we provide a detailed convergence analysis of the numerical methods employed in the main text, including time dependent variational principle numerical simulations, density-matrix renormalization group algorithm, and worldline Monte Carlo method. We extend our investigation to the case of zero interaction between the qubits ($J=0$), demonstrating that the quantum phase transition observed at thermodynamic equilibrium and the dynamical quantum phase transition also occur for a similar critical coupling, which is distinct from the one at non-zero interaction. Furthermore, we explore the manifestations of the quantum phase transition from the perspective of the oscillator by observing its relaxation over time. In the end, we observe that the fidelity of the two qubits has no zeros, implying the absence of dynamical quantum phase transitions. This highlights that only the entire system, not the two-qubits subsystem alone, undergoes a dynamical quantum phase transition.
\end{abstract}
\end{frontmatter}

\newcounter{Cequ}
\newenvironment{Sequation}
  {\stepcounter{Cequ}%
    \addtocounter{equation}{-1}%
    \renewcommand\theequation{S\arabic{Cequ}}\equation}
  {\endequation}

\section{Convergence of numerical methods}
\label{convergence}
\subsection{Time-dependent variational principle numerical simulations}
\label{MPS}
In the main text, we employed time-dependent matrix product state (MPS) simulations, implemented with ITensor Library \cite{fishman2022itensor}, to investigate the system's dynamics, specifically focusing on energy and entanglement behaviors. We analyzed the Loschmidt echo and the corresponding rate function. The long-range interactions between the oscillator, connected to the two qubits, and the bath modes were described using the star geometry. In this configuration, the qubits of frequency $\Delta$ were placed on the first two sites, the oscillator of frequency $\omega_0=\Delta$ and Hilbert space dimension $N_{osc}$ on the third one, and the collection of $N$ bosonic modes of the bath with frequencies $\omega_i$ on the remaining sites. The couplings between the oscillator and each bosonic mode were defined to describe the bath in terms of an Ohmic spectral density, as explained in the main text.\\
The bath Hamiltonian from Eq. (2) in the main text can be expressed as follows:
\begin{equation}
H_{B}=\sum_{i=1}^N \left[\omega_i a^{\dagger}_i a_i +\frac{x_0^2}{2} M_i\omega_i^2\right]- (a+a^{\dagger})\sum_{i=1}^N\left|\lambda_i\right| (a_i+a^{\dagger}_i).
\label{eq:model}
\end{equation}
The coupling constants to the bath are $\left|\lambda_i\right|=\sqrt{\frac{k_i\omega_i}{4m\omega_0}}$. We neglected the energy shift $\sum_{i=1}^N \omega_i/2$, which does not affect the dynamics. Therefore, the Hamiltonian of the system plus the environment can be rewritten by defining a renormalized oscillator frequency $\bar{\omega}_0=\sqrt{\omega_0^2+\sum_{i=1}^N M_i\omega_i^2/m}$. This frequency ensures that the total energy is bounded from below, and the quadratic form is positive definite. In superconducting circuits, this is natural and leads to the quadratic correction of bosonic modes, ensuring that the resonance of the cavity does not change its value irrespective of the dissipation strength. We also define renormalized coupling strengths $\bar{g}=g\sqrt{\frac{\omega_0}{\bar{\omega}_0}}$ between the qubits and the oscillator, and $\left|\bar{g}_i\right|=\sqrt{\frac{k_i\omega_i}{4m\bar{\omega}_0}}$ between the oscillator and each bath bosonic mode. The total Hamiltonian in our MPS simulations is then given by:
\begin{equation}
\label{eq:Hamiltonian}
    H=-\frac{\Delta}{2}(\sigma_x^1 + \sigma_x^2) + \frac{J}{4}\sigma_z^1\sigma_z^2 +\bar{\omega}_0 b^{\dagger}b+\bar{g}(\sigma_z^1 + \sigma_z^2) (b+b^{\dagger})+\sum_{i=1}^N \omega_i a^{\dagger}_i a_i - (b+b^{\dagger})\sum_{i=1}^N \left[\left|\bar{g}_i\right|(a_i+a^{\dagger}_i)\right],
\end{equation}
where $b,(b^{\dagger})$ is the annihilation (creation) operator for the renormalized oscillator with frequency $\bar{\omega}_0$ and coordinates $\Bar{x}_0=\sqrt{\frac{1}{2m\bar{\omega}_0}}(b + b^{\dagger})$ and $\Bar{p}_0=i\sqrt{\frac{m\bar{\omega}_0}{2}}(b^{\dagger}-b)$. The bath is represented by an Ohmic spectral density: $J(\omega) = \sum_{i=1}^N\left|\bar{g}_i\right|^2\delta(\omega-\omega_i)=\frac{\alpha}{2}\omega\Theta(\omega_c-\omega)$, where $\omega_c$ is the cutoff frequency and $\Theta(x)$ is the Heaviside function. The dimensionless parameter $\alpha$ measures the strength of the oscillator-bath coupling.\\
We note that this model can be mapped \cite{zueco2009qubit,de2023signatures,di2023qubit} in such a way that, by including the oscillator as a further bosonic mode of the bath, the qubits are coupled to the $N+1$ bath modes. We can define the couplings $\beta_l$ between each qubit and each bosonic mode of frequency $\hat{\omega}_l$ and hence describe the bath in terms of an effective spectral density:
\begin{equation}
\label{eq:Jeff}
J_{eff}(\omega)=\sum_{l=1}^{N+1}\left|\beta_l\right|^2\delta(\omega-\hat{\omega}_l) \xrightarrow[N\rightarrow\infty]{}\frac{2g^2\omega_0^2\alpha\omega}{(\omega^2-\omega_0^2-h(\omega))^2+(\pi\alpha\omega_0\omega)^2},
\end{equation}
where $h(\omega)=\alpha\omega_0\omega\log\left[\frac{\omega_c+\omega}{\omega_c-\omega}\right]$. The spectral density is Ohmic at low frequencies: $J_{eff}(\omega) \approx \frac{2 g^2 \alpha}{\omega_0^2}\omega$. Therefore, each qubit is coupled to the same oscillator bath through an effective constant proportional to $g^2 \alpha/ \omega_0^2$. This low-frequency behavior of the mapped model suggests the presence of the quantum phase transition (QPT). We studied the system's dynamics for different values of the qubits-oscillator coupling $g$ in the range $[0.0, 0.9]\Delta$, fixing $\alpha=0.1$ and $\omega_c = 30\Delta$.
As detailed in the main text, we selected the ground state of the Hamiltonian $H_S(g=0)$ as the initial state for simulating the system's dynamics. We applied the time-dependent variational principle (TDVP) \cite{haegeman2011time,haegeman2016unifying,paeckel2019time}, where the time-dependent Schrödinger equation is projected onto the tangent space of the MPS manifold with a fixed bond dimension at the current time.\\
In this study, we employed the two-site TDVP (2TDVP as described in \cite{paeckel2019time}), using a second-order integrator with a left-right-left sweeping approach and a half-time step of $dt/2$. This method exhibits a time-step error of $O(dt^3)$, with accuracy controlled by the MPS bond dimension and the threshold to terminate the Krylov series. We halted the Krylov vectors recurrence when the total contribution of two consecutive vectors to the matrix exponential dropped below $10^{-12}$. While more advanced methods, such as basis extension optimization \cite{zhang1998density,brockt2015matrix}, exist, we opted for convergence in the number of Fock states in the cavity ($N_{osc} = 16$) and bath modes ($N = 300$) with Hilbert space dimension ($N_{bos} = 3$). This approach allowed us to find the optimal compromise between the smallest bond dimension and longest simulation times, by converging also over the time interval that we set to $dt\Delta=0.01$.\\
Our truncation error remained below $10^{-13}$ by requiring a maximum bond dimension of $D_{max} = 50$. Simultaneously, this optimal maximum bond dimension enabled us to achieve a final time for our simulations as large as $t_{final}\Delta=50$. In the following figures we set the final time $t_{final}\Delta=40$ because the system has already reached the equilibrium state.

\subsection{Density-matrix renormalization group algorithm}
\label{DMRG}
We conducted an analysis of the equilibrium properties, namely energy and entanglement, for the entire system by computing its ground state through the density-matrix renormalization group (DMRG) algorithm. The results obtained using DMRG were compared with those acquired through the worldline Monte Carlo (WLMC) method. The model parameters were the same used for the TDVP simulations over time.\\
The DMRG algorithm \cite{white1992density,schollwock2005density,schollwock2011density} is an adaptive approach for optimizing an MPS, approximating the dominant eigenvector of a large matrix H, typically assumed to be Hermitian. This algorithm optimizes two neighboring MPS tensors iteratively, combining them into a single tensor for optimization. Techniques such as Lanczos or Davidson are employed for the optimization, followed by factorization using Singular Value Decomposition (SVD) or density matrix decomposition. This process allows us to restore the MPS form and adapt the bond dimension during factorization, preserving the network's structure. \\
In our implementation of DMRG through the ITensor Library's dedicated function \cite{fishman2022itensor}, we could set several accuracy parameters. These include the maximum and minimum bond dimensions of any bond in the MPS, the truncation error cutoff during SVD or density matrix diagonalizations, the maximum number of Davidson iterations in the core DMRG step, and the magnitude of the noise term added to the density matrix to aid convergence. Convergence was achieved by ensuring that the final ground state energy returned after the DMRG calculation was within our specified numerical precision.

\subsection{Worldline Monte Carlo method}
\begin{figure*}
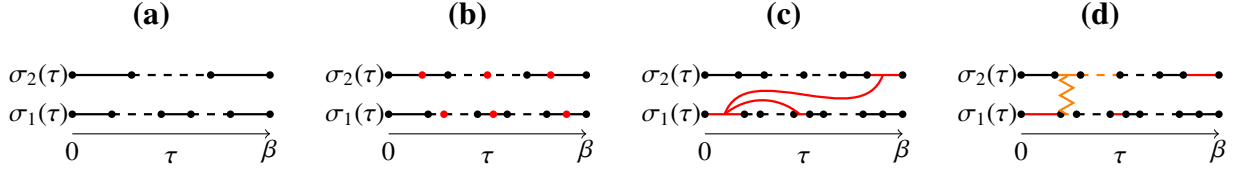

\centering
\begin{tikzpicture}

    \begin{scope}[scale=0.26]
        \input{WLMC-fig/first-step}
    \end{scope}

    \begin{scope}[scale=0.26, xshift=16cm]
        \input{WLMC-fig/second-step}
    \end{scope}

    \begin{scope}[scale=0.26, xshift=32cm]
        \input{WLMC-fig/third-step}
    \end{scope}

    \begin{scope}[scale=0.26, xshift=48cm]
        \input{WLMC-fig/fourth-step}
    \end{scope}
\label{fig:WLMC_scheme}
\end{tikzpicture}
  \caption{(Color online) The scheme of the time cluster algorithm used in this work. In the figure, the dashed lines are the $\sigma(\tau)=-1$ states, and the solid ones are the $\sigma(\tau)=1$ states. The segments are called \textit{super-spin}. (a) Spin path configuration realization; (b) insert randomly, along the spin-paths, $n$ new potential spin flips (in figure are the red dots); (c) connect a random $l$ super-spin to another $m$ with probability $P_{add}(s^I_l,s^{II}_m)$; (d) connect a random $l$ super-spin to another $m$ with probability $P_{add}(s_l,s_m)$.}
  \label{fig:MC-update}
\end{figure*}

WLMC is a path integral technique based on a Monte Carlo algorithm. Using the path integral formulation, it is possible to remove exactly all the phonon degrees of freedom of the thermal bath, obtaining the density matrix dependent only on the effective Euclidean action \cite{winter2009quantum, weiss2012quantum}
\begin{equation}
    S=\frac{1}{2}\sum_{i,j}\int_0^\beta d\tau \int_0^\beta d\tau' \sigma_i(\tau)K_{eff}(\tau-\tau')\sigma_j(\tau'),
\end{equation}
where the effective kernel is
\begin{equation}
K_{eff}(\tau)=\frac{1}{\pi}\int_0^\infty d\omega\ J_{eff}(\omega)\frac{\cosh(\omega(\beta/2-\tau))}{\sinh(\omega\beta\tau)},
\end{equation}
with spectral density
\begin{equation}
J_{eff}(\omega)=\frac{2g^2\omega_0^2\alpha\omega}{(\omega^2-\omega_0^2-h(\omega))^2+(\pi\alpha\omega_0\omega)^2} \Theta(\omega-\omega_c),
\end{equation}
in which $h(\omega)=\alpha\omega_0\omega\log\left[\frac{\omega_c+\omega}{\omega_c-\omega}\right]$. We emphasize that this has the same form as derived in Eq. \ref{eq:Jeff}. Consequently, our simulations are conducted on the mapped system consisting of two qubits interacting with each other and with an effective bath. This scenario resembles that of a quantum Rabi model but is extended to involve two qubits. The problem is therefore equivalent, in the general multi-spin case, to a 2D system in which one dimension is discrete (that of the sites) and one is continuous of length $\beta$. The effective interaction due to $K_{eff}$ consists of a long-range ferromagnetic interaction between the $\tau$ and $\tau'$ points of the worldlines. Between the sites at first neighbours along the discrete dimension (in the case of the studied model) the interaction is antiferromagnetic. The time cluster \cite{winter_2009, rieger1999application} algorithm used (schematised in Figure \ref{fig:WLMC_scheme}) is based on an alternation of Wolff \cite{wolff} and Metropolis moves: in the first step, as schematised in the figure, we start with a worldline $\sigma_z(\tau)$ and add a number of potential spin flips extracted from a Poissonian distribution with mean $\beta\Delta/2$
\begin{equation}
P(n)=\frac{\mu^n}{n!}e^{-\mu},
\end{equation}
where $n=0,1,2,...$ and $\mu=\beta\Delta/2$. Then two of the new segments (having a real spin-flip point and a potential point as extremes) are randomly selected, with extremes $u_1,\ u_2$ and $u_3,\ u_4$. The connection of the two segments occurs with probability
\begin{equation}
    P_{add}(s^I_l,s^{II}_m)=1 - \exp\left\{\min{ \left[ 0, -2\int_{u_1}^{u_2} d\tau\int_{u_3}^{u_4} d\tau's_l^{I}K_\beta(\tau-\tau')s_m^{II} \right] }\right\}.
\end{equation}
This step is iterated connecting the second segment to another randomly selected one. The process is iterated until no more segments are added to the cluster. The cluster is subsequently flipped with probability $1/2$ and the potential spin flips introduced at the beginning that do not represent real flips are removed. The same procedure is carried out for the antiferromagnetic interaction at first neighbours by connecting segments along worldlines of opposite sign, of extremes $u_1',\ u_2'$ and $u_3',\ u_4'$, with probability
\begin{equation}
    P_{add}(s_l,s_m)=1-\exp\left\{\min[0, 4\beta J s_l s_m]\right\}.
\end{equation}
The rest of the algorithm is similar to the previous case. Finally, a Metropolis step is performed in which a segment is randomly selected and flipped, thus removing the extreme points that define the spin flips. We emphasise that this approach is exact from a numerical point of view, and it is equivalent to the sum of all the Feynman diagrams.

\section{Non-interacting qubits case (J=0)}
\label{J0}
\begin{figure}[!htbp]
    \begin{center}
        \includegraphics[scale=0.4]{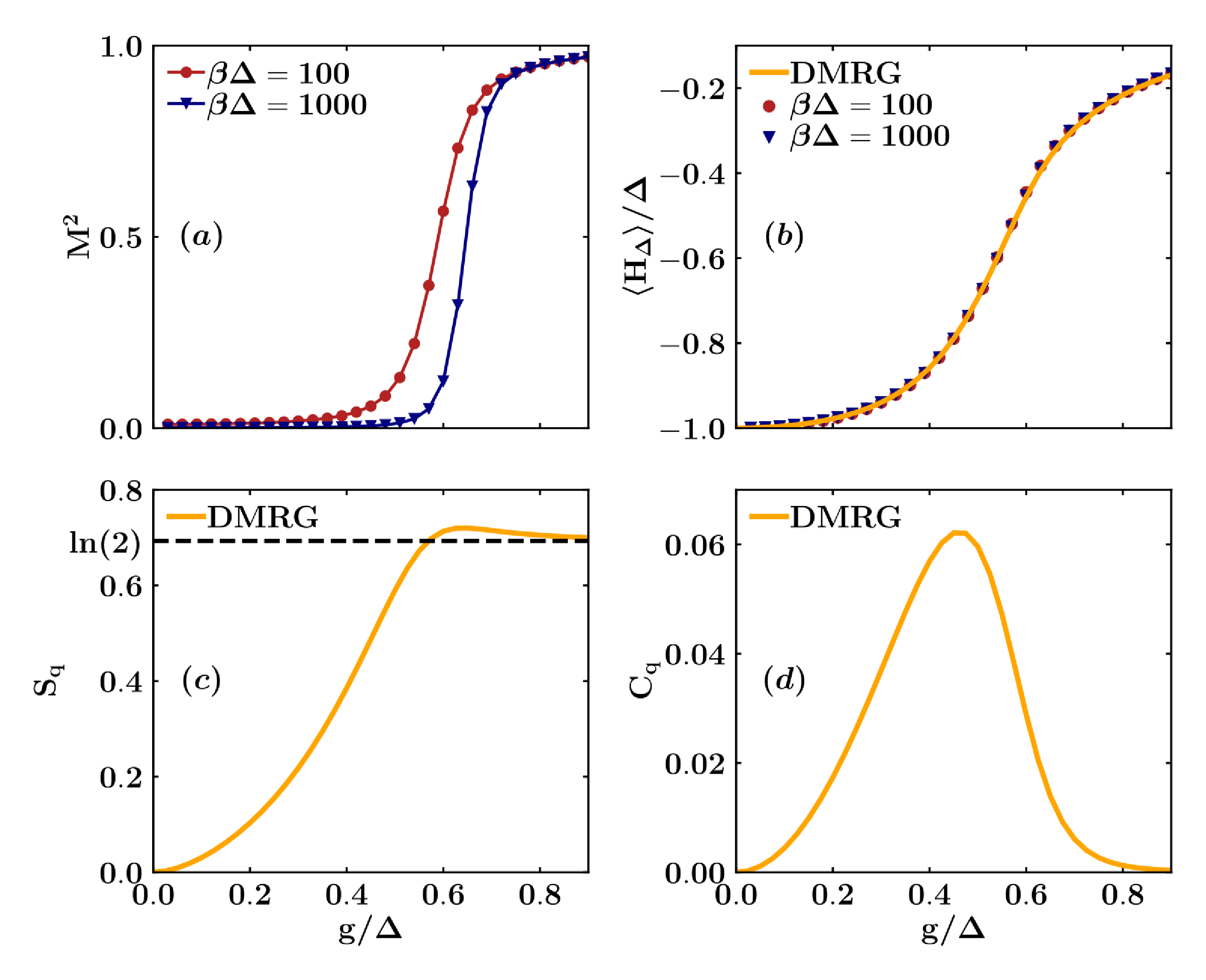}
     
        \caption{\label{fig:te1sm}(Color online) Qubits' squared magnetization $M^2$ ($a$), qubits' energy $\langle H_{\Delta}\rangle/\Delta$ ($b$), entropy $S_q$ ($c$) and concurrence $C_q$ ($d$) as functions of $g/\Delta$, computed through WLMC and DMRG. For the WLMC method the calculations are made for $\beta\Delta \in[100,1000]$.} 
    \end{center}
\end{figure}

In this section, we explore the case where the interaction ($J$) between the two qubits is zero. Here, we demonstrate the agreement in thermodynamic quantities computed through both DMRG and WLMC methods. Further, we have estimated the critical coupling ($g_c\approx0.6\Delta$) using an analysis of the Minnhagen function. Figure \ref{fig:te1sm}.a presents the squared magnetization of the qubits ($M^2$), plotted as a function of $g/\Delta$ for two different inverse temperatures, $\beta\Delta=100$ and $\beta\Delta=1000$. With the WLMC method, we observe that the squared magnetization changes sharply from $0$ to $1$ at a critical value of $g$. This jump becomes more pronounced as the temperature decreases. Furthermore, we calculate the mean value of the qubits Hamiltonian, denoted as $\langle H_{\Delta}\rangle = -\Delta\langle\sigma_x^1+\sigma_x^2\rangle/2$, as a function of $g/\Delta$ for the same two temperature values (see Figure \ref{fig:te1sm}.b). Both the WLMC and DMRG methods yield consistent results. The qubits Hamiltonian does not exhibit a jump but becomes less negative as the bath reduces the effective qubits gap with increasing values of $g$.
\begin{figure}[!htbp]
    \begin{center}
        \includegraphics[scale=0.4]{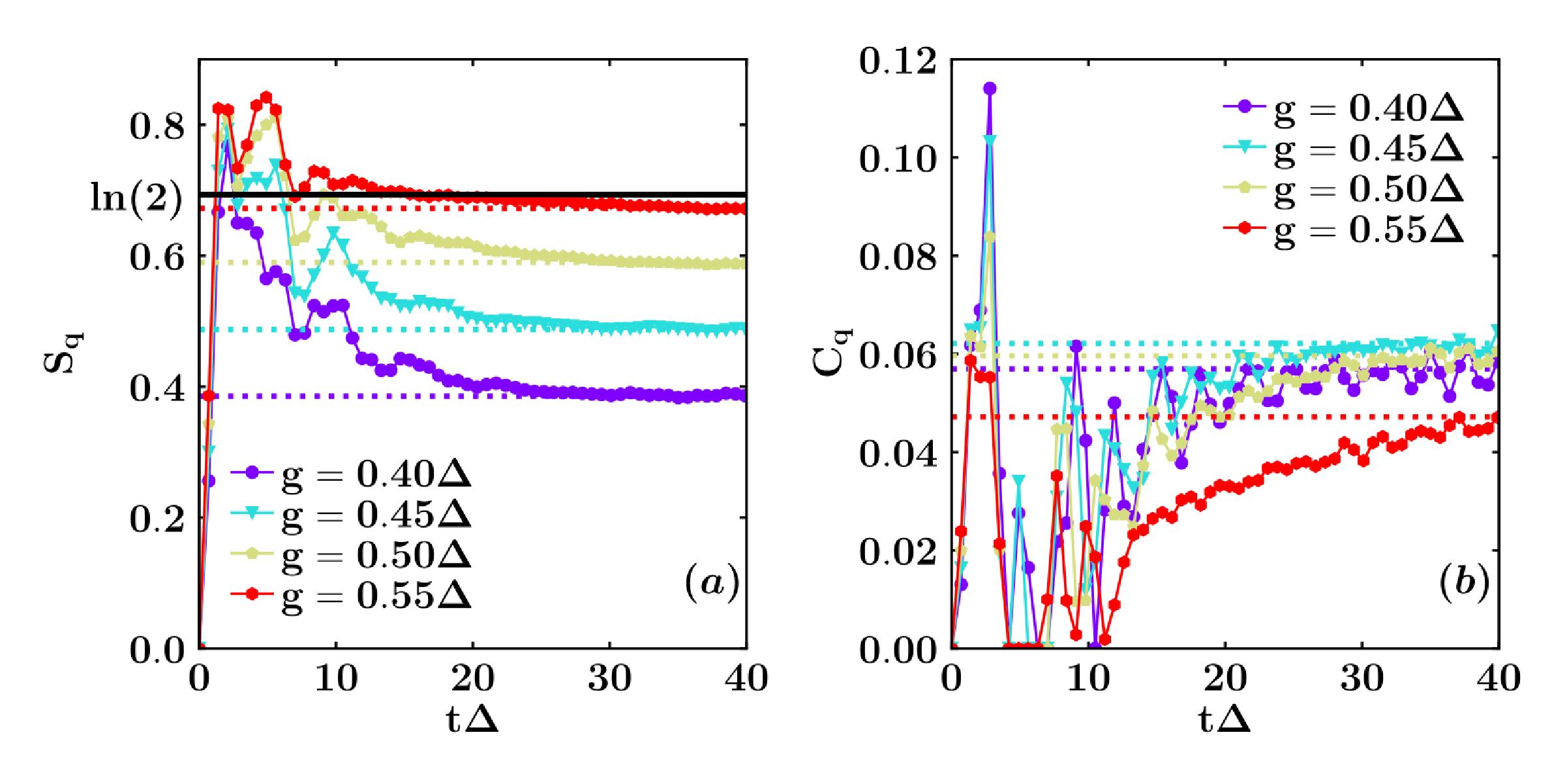}
     
        \caption{\label{fig:enentsm}(Color online) Qubits' entropy $S_q$ ($a$) and concurrence $C_q$ ($b$) as functions of dimensionless time $t\Delta $ for different values of the coupling $g\in[0.40,0.55]\Delta$, near the critical point, computed through TDVP and DMRG.} 
    \end{center}
\end{figure}
To analyze the entanglement properties of the system in the presence of the QPT, we compute the qubits' entropy, denoted as $S_{q}$, and the concurrence, given by $C_{q}$. Figures \ref{fig:te1sm}.c and \ref{fig:te1sm}.d present the entropy and entanglement as functions of $g/\Delta$, computed using the DMRG algorithm. The entropy increases for values of $g$ near the critical point, asymptotically approaching a value of approximately $\ln(2)$. Conversely, the concurrence is almost zero everywhere, except for a small peak near the transition. This behavior can be explained by the qubits approaching a two-degenerate state during the Beretzinski-Kosterlitz-Thouless (BKT) QPT, where they are either both in the ``up" or ``down" state, resulting in a lack of entanglement and an effective two-state density matrix. The difference from the case of non-zero $J$ is that the concurrence does not change much because the qubits prefer to entangle with the bath to facilitate the transition and entropy can be slightly greater than $\ln(2)$. \\

\begin{figure}[!htbp]
    \begin{center}
        \includegraphics[scale=0.36]{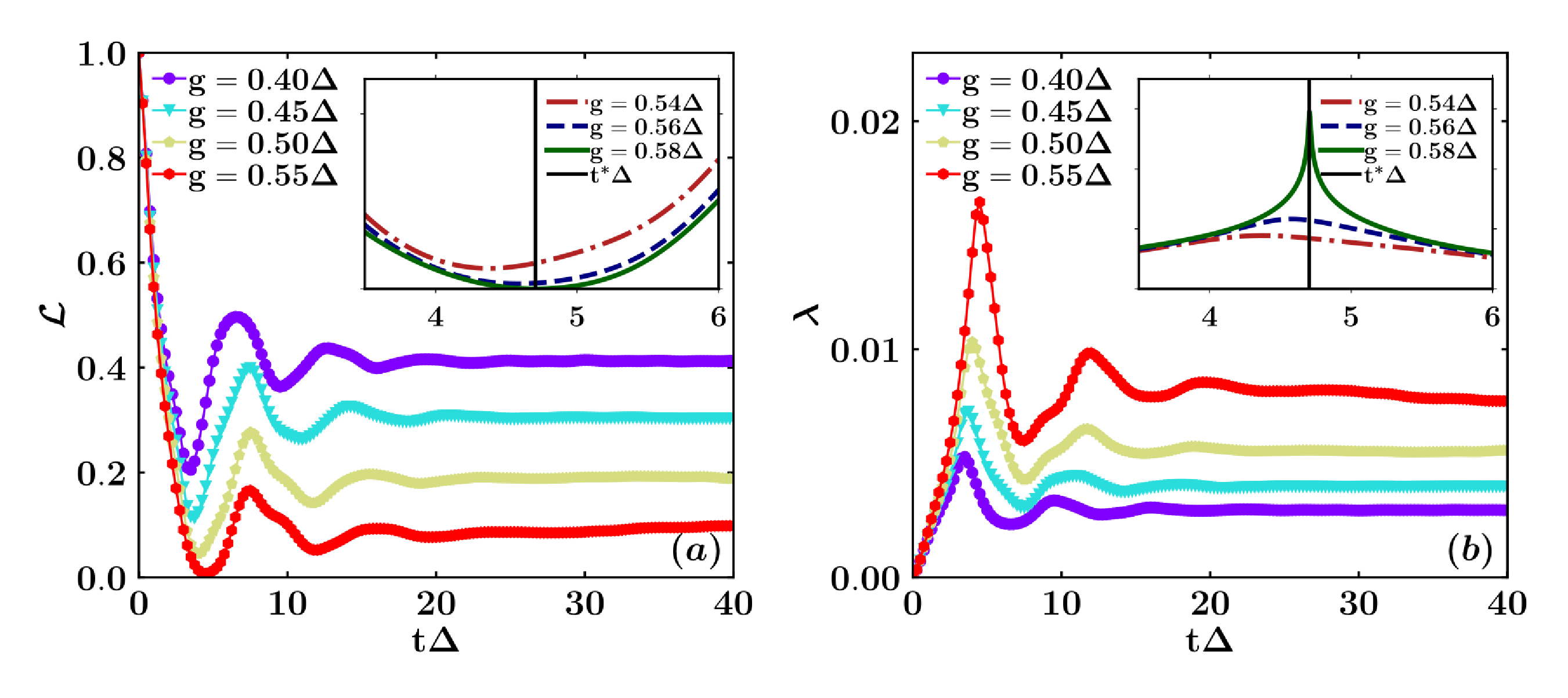}
     
        \caption{\label{fig:echosm}(Color online) Loschmidt echo $\mathcal{L}(t)$ ($a$) and rate function $\lambda(t)$ ($b$) as functions of dimensionless time $t\Delta $ for different values of the coupling $g\in[0.40,0.55]\Delta$, near the critical point, computed through TDVP. The insets provide a zoomed-in view near the transition allowing to identify the critical time $t^*\Delta\approx4.7$.} 
    \end{center}
\end{figure}

We then investigate the out-of-equilibrium properties of the system, focusing on the entanglement over time. To accomplish this, we employ the TDVP algorithm. Additionally, we compare these behaviors at very long times with those computed using the DMRG at thermodynamic equilibrium.\\
Specifically, we calculate the qubits' entropy, denoted as $S_q(t)$, and the concurrence $C_q(t)$ for different values of $g$, approaching the critical point. Figures \ref{fig:enentsm}.a and \ref{fig:enentsm}.b demonstrate that both entropy and entanglement approach thermodynamic values at earlier times than the $J\neq 0$ case. Moreover, the greater the value of $g$, the more time the system needs to reach equilibrium values.\\
We do not show the time behavior of the energy contributions, but we have analyzed them, finding results similar to the interacting case. That is, the system's energy and the interaction with the bath approach equilibrium values, while the bath's energy remains different, accounting for the overall difference in energy due to the initial excited state.\\
Finally, we focus on the occurrence of the dynamical quantum phase transition (DQPT) in our model by computing the Loschmidt echo and the corresponding rate function over time. Figure \ref{fig:echosm} illustrates the echo and the rate function over time for different values of $g$ around the transition. Additionally, the inset in Figure \ref{fig:echosm}.a clearly demonstrates how the scalar product between the evolved state and the initial one becomes zero as the kink becomes narrower and higher, especially as the critical point is approached at time $t^*\Delta \approx 4.7$ (see inset of Figure \ref{fig:echosm}.b). Beyond the critical value $g_c\approx 0.6\Delta$, consistent with the one obtained through the Minnhagen function analysis, the peak occurs at earlier times, and multiple peaks emerge over time in the rate function. As in the interacting case, fluctuations in energy could be responsible for the observation of the DQPT in the excited states, reminiscent of the QPT occurring at thermodynamic equilibrium at zero temperature in the ground state of the entire Hamiltonian. This observation of the transition even when going out of equilibrium suggests that we are exploring the first excited states of the Hamiltonian. We have also conducted the same scaling analysis in terms of $N$ and verified that at the critical coupling, the rate function remains finite as $N$ approaches infinity, thus demonstrating this behavior in the thermodynamic limit.
\section{QPT evidences on the oscillator in the non-interacting qubits case (J=0)}
\label{oscillator}
\begin{figure}[!htbp]
    \begin{center}
        \includegraphics[scale=0.45]{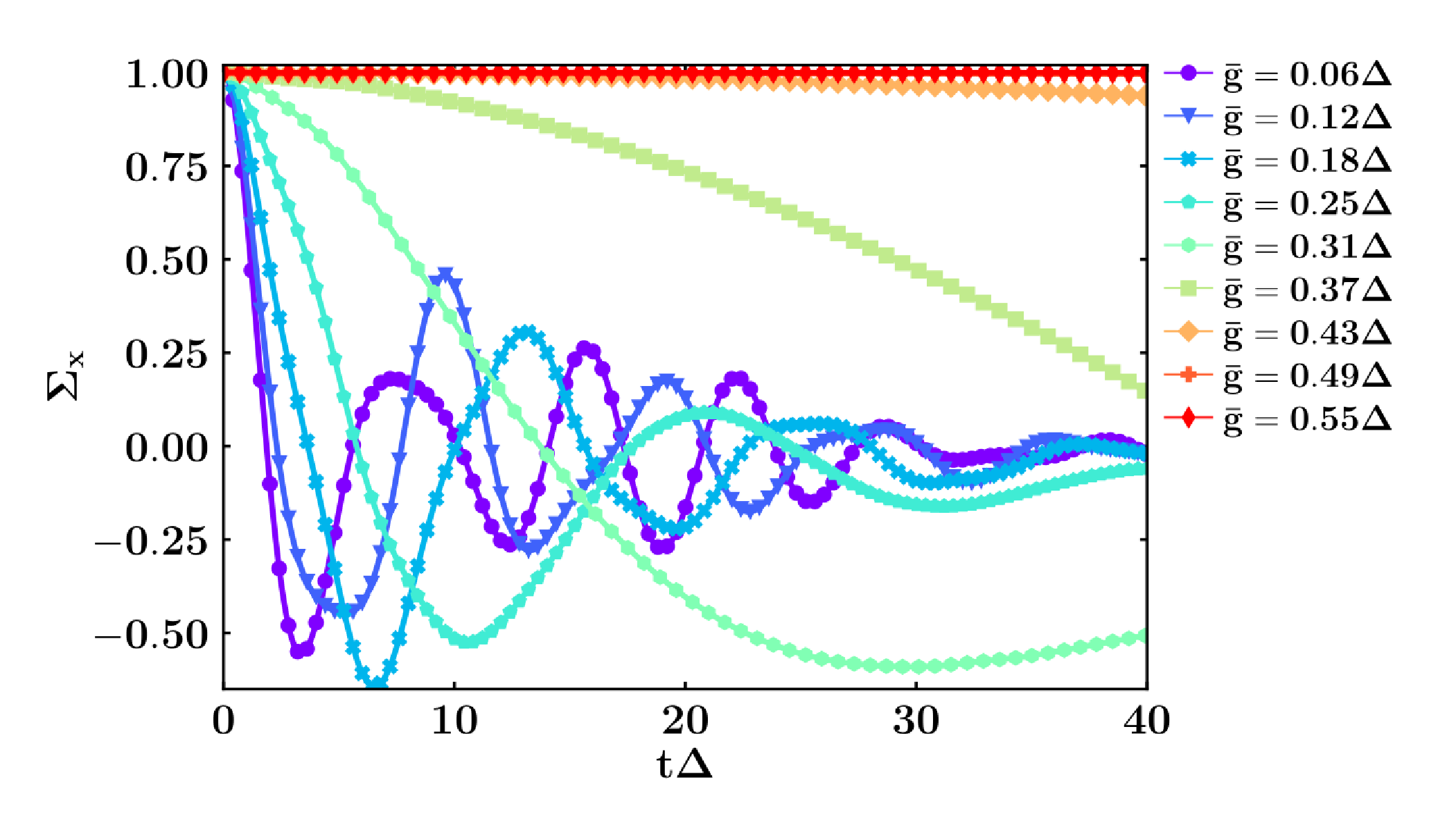}
     
        \caption{\label{fig:relaxsm}(Color online) Oscillator's relaxation function $\Sigma_x(t)$ as a function of dimensionless time $t\Delta $ for different values of the renormalized coupling $\Tilde{g}\in[0.06,0.55]\Delta$, near the critical point, computed through TDVP, by adding a small magnetic field on the qubits $\epsilon=0.001\Delta$. } 
    \end{center}
\end{figure}

In the following, we investigate the physical features of the relaxation function involving the oscillator position operator for the non-interacting two-qubits case. We adopt the star geometry again to describe the long-range interactions between the oscillator and the bath modes, using the same Hamiltonian form as in Subsection \ref{MPS} with identical parameters. To study the relaxation of the oscillator, suppose that the system at $t = -\infty$ is at thermal equilibrium. The response of the system to a perturbation, adiabatically applied from $t = -\infty$ and cut off at $t = 0$, can be calculated within the Mori formalism and the linear response theory \cite{de2023signatures}.\\
In particular, we adiabatically apply a small electric field acting on the renormalized oscillator, defined as $H_{field}=\epsilon b^{\dagger}b$ with $\epsilon=10^{-3}$. One crucial physical quantity for all $t \geq 0$ is the relaxation function $\Sigma_{x}(t) = \frac{(x(t);x(0))}{(x(0);x(0))} = \frac{\langle x(t)\rangle}{\langle x(0)\rangle}$ (calculated in the absence of $\epsilon$, where $t\geq 0$). In Figure \ref{fig:relaxsm}, we plot $\Sigma_{x}(t)$. Even at $g = 0$, the Rabi oscillations of the oscillator are already damped due to the interaction with the environment. The amplitude and frequency of these oscillations further reduce with an increase in the coupling strength $g$. Subsequently, the relaxation becomes exponential, with the relaxation time progressively increasing. At $g\geq g_c$, the system does not reach a relaxed state ($\Sigma_{x}(t) = 1$) independently of time $t$, signaling the occurrence of a QPT.\\
Once again, the ground state with the electric field is computed using the DMRG algorithm. We employ the 2TDVP, as before, to compute the relaxation over time.

\section{Fidelity of the two qubits}
\begin{figure}[!htbp]
    \begin{center}
        \includegraphics[scale=0.45]{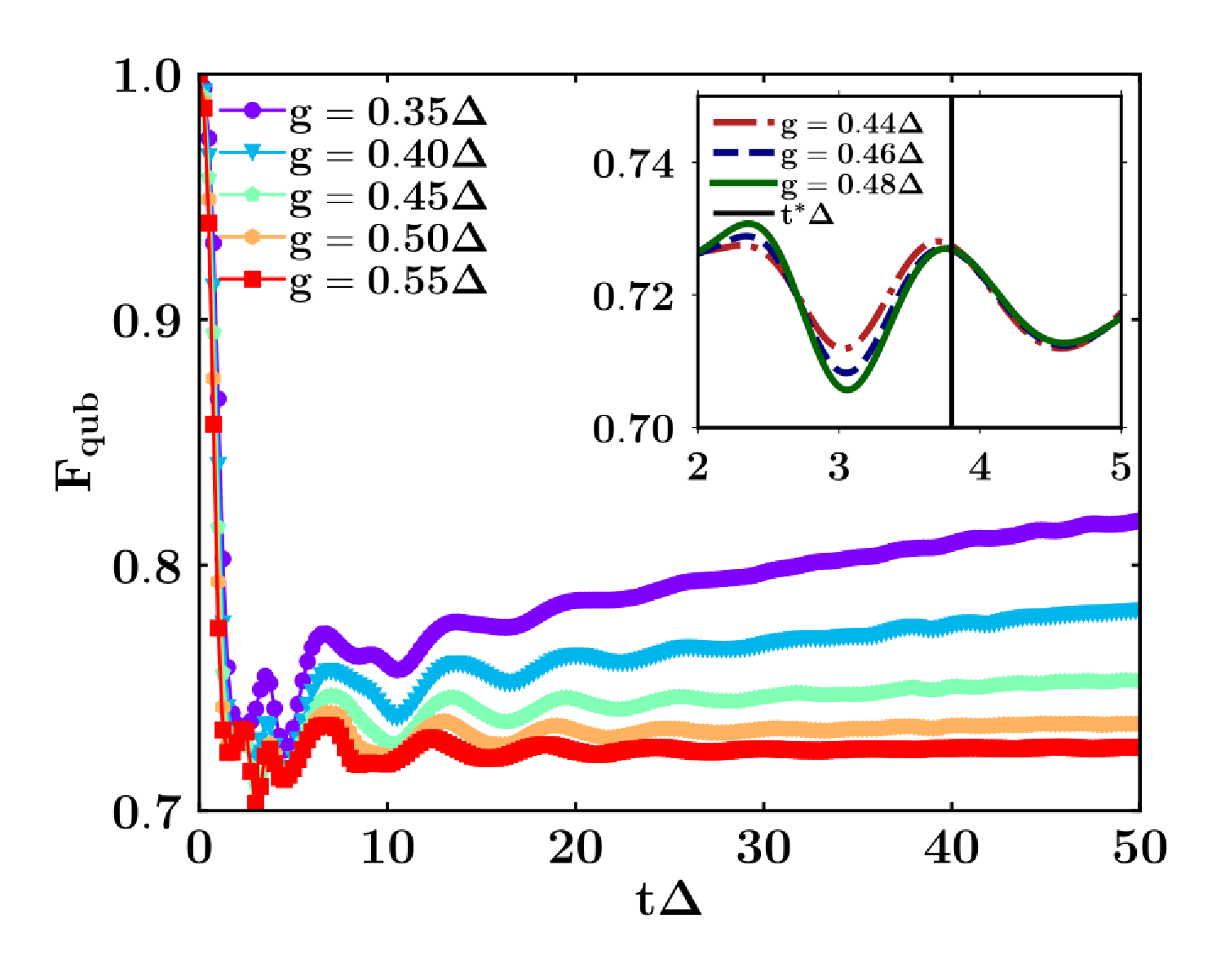}
     
        \caption{\label{fig:fidelitysm}(Color online) Qubits' fidelity $F_{qub}(t)$ as a function of dimensionless time $t\Delta $ for different values of the coupling $g\in[0.35,0.55]\Delta$, crossing the critical point, computed through TDVP. } 
    \end{center}
\end{figure}

Here, our focus is on examining the fidelity of the two qubits to identify potential zeros, akin to the Loschmidt echo. Specifically, the absence of zeros indicates the absence of a dynamical phase transition in the qubits' system. We calculate the time evolution of the fidelity, defined as follows:
\begin{equation}
    F_{qub}(t)\equiv F(\rho_{qub}(0),\rho_{qub}(t))=\Tr \sqrt{\rho_{qub}(0)^{1/2}\rho_{qub}(t)\rho_{qub}(0)^{1/2}}.
\end{equation}
Here, $\rho_{qub}$ represents the reduced density matrix of the two qubits. This computation involves comparing the evolved state of the qubits at time $t$ with their initial state, similar to the Loschmidt echo for the entire system.\\
Figure \ref{fig:fidelitysm} illustrates how, for $J=-10\Delta$ (as in the main text) and various values of $g$, even those very close to the critical point (refer to the inset of Figure \ref{fig:fidelitysm}), the fidelity never reaches zero. Instead, it consistently remains around $0.7$, even as $g$ increases.

\bibliographystyle{apsrev4-2}

\bibliography{biblio}